\documentstyle[epsfig, twocolumn]{esapub}

\newcommand{\lesssim}{\mathrel{\hbox{\rlap{\hbox{\lower4pt\hbox{$\sim$}}}\hbox{$<$}}}}
\newcommand{\gtrsim}{\mathrel{\hbox{\rlap{\hbox{\lower4pt\hbox{$\sim$}}}\hbox{$>$}}}}

\newcommand{\et}{\it et~al.}                  %

\begin{document}

\setlength{\parindent}{0pt}
\setlength{\parskip}{10pt plus 1pt minus 1pt}
\setlength{\hoffset}{-1.5truecm}
\setlength{\textwidth}{17.1 true cm}
\setlength{\columnsep}{1truecm}
\setlength{\columnseprule}{0pt}
\setlength{\headheight}{12pt}
\setlength{\headsep}{20pt}
\pagestyle{esapubheadings}

\title{\bf ISOCAM DEEP SURVEYS UNVEILING STAR FORMATION IN THE MID-INFRARED}

\author{\bf D.~Elbaz$^1$, H.~Aussel$^1$, A. C. Baker $^1$, C.J.~C\'esarsky$^1$, D. L. Clements$^2$, F.X.~D\'esert$^2$, D.~Fadda$^1$,\\
\bf A.~Franceschini$^3$, J.L.~Puget$^2$, J.L.~Starck$^1$
\vspace{2 mm} \\ 
$^1$Service d'Astrophysique, CEA/DSM/DAPNIA Saclay,
Orme des Merisisers, 91191 Gif-sur-Yvette C\'edex, France \\ fax
$+$33.1.69.08.65.77, e-mail {\tt elbaz@cea.fr} \\ 
$^2$Institut d'Astrophysique Spatiale, B{\^a}t 121, Universit\'e Paris XI, F-91405 Orsay C\'edex, France \\ 
$^3$Osservatorio Astronomico di Padova, Italy}

\maketitle


\begin{abstract}
	
	Before having exhausted the helium in its tank on April 8,
1998, the Infrared Space Observatory had time to perform several
complementary surveys at various depths and areas within selected
regions of the sky. We present the results of some of the surveys done
with the mid-infrared camera, ISOCAM, on-board ISO, and more
specifically those surveys which were performed using the broad-band
filter LW3 (12-18 $\mu$m).

The counts obtained above 1 mJy by various surveys in the 12-18 $\mu$m
band are in good agreement with the no-evolution model. On the
contrary, a strong evolution is observed in the counts below 1 mJy by
deeper surveys. This evolution seems to be due to a population of
star-forming or post starburst galaxies, whose optical properties are
not very different from the galaxies in the field. This implies that
most of the star formation is hidden by dust and appears in the
infrared as redshift rises.
\vspace{5 pt} \\

Keywords: cosmological surveys; mid-infrared; galaxy evolution.
\end{abstract}


\section{INTRODUCTION}
\label{SECTION:intro}

	The so-called 'Madau plot' (Madau $\et$ 1996) which follows
the rate of star formation per unit comoving volume in the universe as
a function of redshift presents the advantage of being global
enough to give an indication of the past history of star formation as
a whole, avoiding the difficulties encountered when trying to separate
the respective role of individual galaxy types (Sp/Pec/E, hierarchical
galaxy formation {\it vs.} individual collapse, duration of the
bursts). However, based on optical-UV information only, it is now clear
that this plot only sets a lower limit to the total star formation
history of the universe (see below).

This plot also addresses the question of the heavy element content of
the universe as a way to normalize the past history of star formation
in the universe. Assuming that there was little margin left for more
star formation than already found from optical-UV light one could see
the lower limits given by Madau $\et$ 1996 as if they were firm
constraints. Since then Lyman-break galaxies (also called UV drop-out
galaxies) found by Steidel $\et$ (1996) appeared to be strongly
affected by dust obscuration. Pettini $\et$ (1997) estimated that
about two thirds of the emission due to star formation taking place in
these galaxies is radiated in the infrared as indicated by the slope
of their rest-frame UV spectral energy distribution (see also Meurer
$\et$ 1997, for the strong UV absorption at high redshifts).  The
metal content of the local universe presents enough uncertainties to
leave margin for hidden star formation, especially at high redshifts
where time flows faster per unit redshift. Among those uncertainties,
we shall quote the lack of galactic globular clusters with more than
solar metallicity which would be required to calibrate the metallicity
of metal-rich stellar populations found in elliptical galaxies; the
presence of at least as much metals in the intra-cluster medium of
galaxy clusters than in the galaxies themselves (Arnaud $\it{et~al.}$
1992, Mushotzky $\&$ Loewenstein 1997), with an over-abundance of
$\alpha$-elements which may indicate a biased IMF in the early stages
of star formation in ellipticals (Elbaz $\it{et~al.}$ 1995); the still
uncertain value and universality of the IMF. There is clearly space
for a factor 2-4 in the heavy elements content of the local universe
which implies a large freedom in the star formation history of the
distant universe where time is so short.

Although these indirect arguments already favour dust-enshrouded star
formation with respect to optical-UV estimates in distant galaxies,
the correction for dust obscuration remains controversial and direct
observations in the infrared are critical at this stage. The obvious
way to go for it is the far infrared (FIR) since the spectral energy
distribution (SED) of dusty star forming regions peaks around 60
$\mu$m. IRAS was not sensitive enough to probe the universe above
z$\simeq$0.2, except for some exceptionnally bright objects such as
10214$+$4724 (Rowan-Robinson $\et$ 1991). FIRAS on-board COBE detected
a diffuse far-infrared to sub-mm background (Puget $\it{et~al.}$ 1996,
Guiderdoni $\it{et~al.}$ 1997, Fixsen $\it{et~al.}$ 1998). The origin
of this Cosmic InfraRed Background (CIRB) remains controversial and
could either be due to a local population of galaxies (z $\simeq$ 1)
as proposed by Fall, Charlot $\&$ Pei (1996) or by a population of
galaxies distributed over a wider redshift range as suggested by Puget
$\et$ (1996 and this conference, and Lagache $\it{et~al.}$ 1998).  The
recent detection of a population of distant dusty galaxies at 450 and
850 $\mu$m with the sub-mm array camera SCUBA on the JCMT (Smail,
Ivison $\&$ Blain 1997, Blain $\et$ 1998, Cimatti $\et$ 1998, Barger
$\et$ 1998a, Hughes $\et$ 1998) tends to favour a shift of the peak of
the star formation rate per unit comoving volume towards larger
redshifts. These observations would favour the distant galaxy
population option to explain the CIRB. However, finding counterparts
with spectroscopic redshift to these SCUBA galaxies will be the next
step to answer this question. There will also still remain an
uncertainty on the FIR part of the SED since sub-mm wavelengths only
probe the right side of the peak, even at z=3 where the peak of the
SED is redshifted to 240 $\mu$m. Reliable quantitative estimates of
the star formation rate (hereafter SFR) will only become available
when the rest-frame FIR part of the SED will be sampled. For this
purpose, ISOPHOT on-board ISO, is not sensitive enough, although
tantalizing results are already obtained at 175 $\mu$m (Kawara $\et$
1998, Lagache $\et$ 1998, Puget $\et$ 1998). The satellite FIRST will
certainly play a major role in this respect.

However, one can already see that hidden star formation plays a
major role. Although the mid-infrared range is less directly
correlated with star formation than larger wavelengths, we will show
that with a 1000 times better sensitivity and 60 times better spatial
resolution than IRAS, ISOCAM mid-infrared deep surveys confirm that
most star formation is hidden by dust and that this effect increases
with redshift ($z<1.5$).


\section{ORIGIN OF THE MID-INFRARED EMISSION AND ITS LINK WITH STAR FORMATION}
\label{SECTION:origin}

\begin{figure}[!ht]
\centerline{\psfig{file=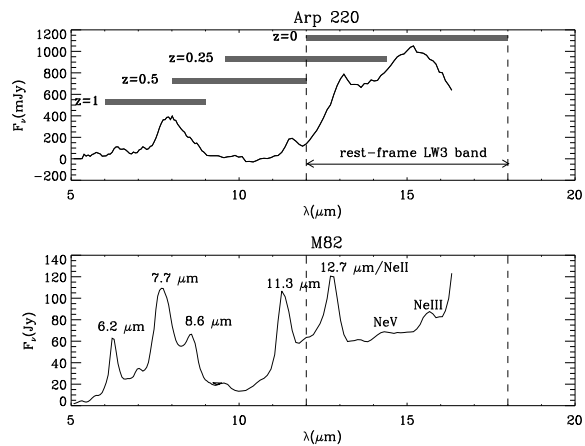,height=6.0cm,width=8.0cm}}
\caption{\em SEDs from ISOCAM Circular Variable Filter. Upper plot:
Arp 220 (Charmandaris $\et$ 1998), UIBs nearly absent, high ratio of
the 12-18 $\mu$m over the 5-8.5 $\mu$m band. The displacement of the
12-18 $\mu$m (LW3) bandpass with redshift has also been plotted. Lower
plot: M82 (Tran 1998), strong UIB features and low ratio of the 12-18
$\mu$m over the 5-8.5 $\mu$m band.}
\label{FIGURE:cvf}
\end{figure}

The rest frame mid-IR emission of galaxies can be divided into three
components (Puget $\&$ L\'eger 1989, D\'esert $\et$ 1990):

\begin{itemize}
\item
{\bf UIBs:} the large Unidentified Infrared Bands (UIBs), detected at
6.2, 7.7, 8.6, 11.3 and 12.7 $\mu$m as well as their underlying
continuum, dominate the mid-IR emission below 12 $\mu$m (see
figure~\ref{FIGURE:cvf}).  The carriers of these UIBs contain
Polycylic Aromatic Hydrocarbons (PAHs, L\'eger $\&$ Puget 1984, Puget
$\&$ L\'eger 1989). The fact that in the wavelength range where the
emission is dominated by UIBs, it globally scales with the intensity of
the radiation field with only small changes in the spectrum shape is a
strong indication that the carriers of the UIBs are transiently heated
by the absorption of individual photons (Boulanger 1998). The
mid-infrared emission of galaxies does increase with the star
formation rate though saturating for strong radiation fields (Boselli
$\et$ 1998).
\item 
{\bf Warm dust (T$>$150 K):} continuum at {\bf $\lambda > 10~\mu m$}
from Very Small Grains (VSGs) of dust (D\'esert $\et$ 1990).
\item
{\bf Forbidden lines of ionized gas:} NeII (12.8 $\mu m$), NeIII (15.6
$\mu m$), SIV (10.5 $\mu m$), ArII (7 $\mu m$). These lines are good
indicators of the star formation activity, particularly the NeIII/NeII
ratio, when using the low resolution spectro-imagery mode of ISOCAM
for nearby galaxies, but their contribution to large bands such as
those used for the deep surveys is negligible.
\end{itemize}

All three components affect the 12-18 $\mu$m ISOCAM filter (LW3) due
to K-correction for galaxies between redshifts 0 and 1.5. Indeed, the
LW3 band includes emission from very small grains for low redshifts
and becomes more and more contaminated by UIBs with increasing
redshift.

The Far Infrared (FIR) flux is usually assumed to be correlated to the
SFR once the cirrus emission is removed (see Scoville $\&$ Soifer
1991, Kennicutt 1998). The MIR emission is also a strong indicator of
star formation, but the derivation of a star formation rate
exclusively from the MIR flux is tricky: while many active star
forming galaxies exhibit comparable infrared SEDs, some are very
different. The mid-IR over far-IR ratio can vary from less than 1
$\%$ to a few times 10 $\%$. This results directly from the complex
underlying physics of the mid-infrared emission which differs from the
FIR one.  For example, let us compare Arp 220 and M82 (see
table~\ref{TABLE:compir} and figure~\ref{FIGURE:cvf}):

\begin{table}[htb]
\begin{center}
\caption{\em Comparison of M82 \& Arp 220: two template nearby
starburst galaxies (H$_{o}$=75 km.s$^{-1}$.Mpc$^{-1}$, distances
corrected for the Virgo infall: d(M82)=3 Mpc, d(Arp220)= 73 Mpc). See
Genzel $\et$ (1998) for a discussion showing that the emission of Arp
220 is dominated by its starburst activity and not its central AGN.}
\vspace{0.5 em}
\begin{tabular}{lccc}
\hline \\[-5 pt]
Waveband & Arp 220 & M82 & $\frac{Arp220}{M82}$\\
\hline \\[-5 pt]
Visible (L$_{V}^{Tot}$) & 2.6$\times10^{9}~L_{\odot}$	& 4.3$\times10^{8}~L_{\odot}$	& 6 \\
L$_{LW3}$ (12-18 $\mu$m) & 7.5$\times10^{9}~L_{\odot}$ & 1.6$\times10^{9}~L_{\odot}$& 4.7 \\
L$_{FIR}$ (8-1000 $\mu$m) &  1.3$\times10^{12}~L_{\odot}$ & 2.5$\times10^{10}~L_{\odot}$& 52 \\
$L_{LW3}/L_{V}^{Tot}$ &  2.9 & 3.7 & 1.3 \\
$L_{FIR}/L_{V}^{Tot}$ &  500 & 58 & 8.6 \\
$L_{FIR}/L_{LW3}$ &  173 & 16 & 11 \\
\hline
\end{tabular}
\label{TABLE:compir}
\end{center}
\end{table}

Both galaxies emit approximately the same 15 $\mu$m flux relative to
visible light (respectively 3 , for Arp 220, and 4, for M82, times
more energy is radiated at 15 $\mu$m than at 0.55 $\mu$m). The
contribution of the far infrared part of the spectrum relative to the
MIR is substantially different: the ratio of the FIR to MIR light is
10 times larger for Arp 220 than for M82. This difference of an order
of magnitude in FIR luminosity would imply one order of magnitude
difference in star formation rate (using the formula given by Scoville
$\&$ Soifer 1991 or Kennicutt 1998). Hence, the mid-IR flux cannot be
considered as a direct measure of the star formation rate. However,
Arp 220 is an extreme case and most galaxies behave more like M82 than
Arp 220, so that the uncertainty is probably closer to 5 than to 10.

\begin{figure}[!ht]
\centerline{\psfig{file=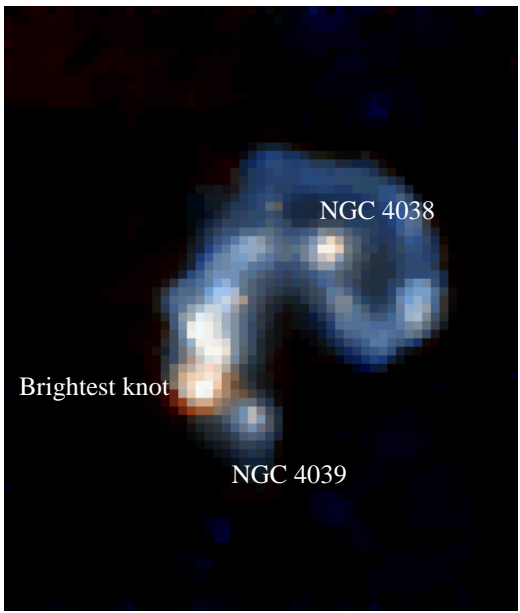,height=8.0cm,bbllx=1,bblly=50,bburx=317,bbury=419}}
\caption{\em mid-infrared image of the Antennae galaxies. The brightest
spot lies in the overlapping region of the two interacting galaxies
NGC 4038/4039, while the optical-UV emission peaks at the position of
the two galactic nuclei (Vigroux et al. 1996)}
\label{IMAGE:antennae}
\end{figure}

If the link between mid-IR fluxes and star formation is not simple and
implies the need for a multi-wavelength approach to reach reliable
quantitative estimates of the star formation rate, the case of the
Antennae galaxy is sufficiently explicit by itself to prove that
hidden star formation can be unveiled through its mid-IR emission
(Mirabel $\et$ 1998). ISOCAM mid-infrared imaging indeed revealed a
bright spot (see figure~\ref{IMAGE:antennae}, from Vigroux $\et$ 1996)
which emits alone 15 $\%$ of the total mid-IR emission, within the
overlapping region of the two interacting galaxies NGC 4038/4039
(Mirabel $\et$ 1998). This spot is clearly optically obscured and most
of the optical emission arises from the two galactic nuclei.

\begin{figure}[!ht]
\centerline{\psfig{file=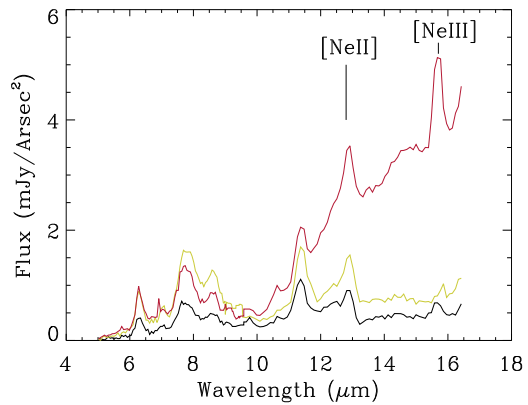,height=6.0cm,width=8.0cm}}
\caption{\em CVF spectra of 3 regions in the Antennae galaxy (NGC
4038/4039): the 2 bottom plots correspond to the 2 galaxy nuclei,
while the upper curve is associated to the brightest 15 $\mu$m spot in
the overlapping region of the 2 galaxies.}
\label{FIGURE:spec_ant}
\end{figure}

The spectral energy distribution at the position of this spot presents
two emission lines, NeII and NeIII, which do not appear as clearly on
the brightest optical spots corresponding to the two galaxy nuclei,
with a ratio of NeIII over NeII close to one, hence indicating the
presence of massive young stars (figure~\ref{FIGURE:spec_ant}, Mirabel $\et$
1998).

\section{ISOCAM DEEP SURVEYS}
\label{SECTION:ds}

\begin{figure}[!ht]
\centerline{\psfig{file=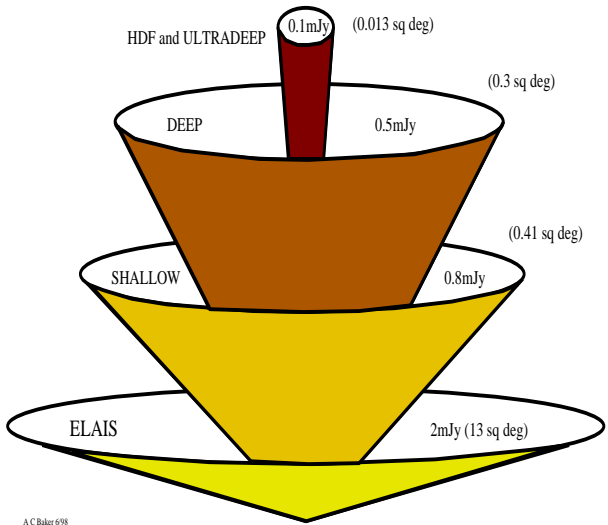,height=8.0cm,width=8.0cm}}
\caption{\em ISOCAM extragalactic surveys.}
\label{FIGURE:cones}
\end{figure}

Several extragalactic surveys have been performed using ISOCAM on-board ISO
ranging from very large and shallow ones (ELAIS,
P.I. M. Rowan-Robinson) to smaller and deeper ones
(figure~\ref{FIGURE:cones}).

As shown by the number counts plot (section~\ref{SECTION:results}),
all these surveys are complementary since they explore the behaviour
of galaxies in the mid-IR over nearly 4 orders of magnitude in the Log
N-Log S plot. All surveys have been performed in regions of the sky
relatively clean of cirrus and zodiacal emission: the Lockman Hole and
Marano Field regions, but also the Hubble Deep Field region. The main
limitation of these surveys comes from the presence of cosmic ray
impacts on the detector for the detection and completeness limits (see
below) and from the transient behaviour due to the very low
temperature of the detectors for the photometric accuracy.

\subsection{Data reduction}
\label{SECTION:data}

\begin{figure}[!ht]
\centerline{\psfig{file=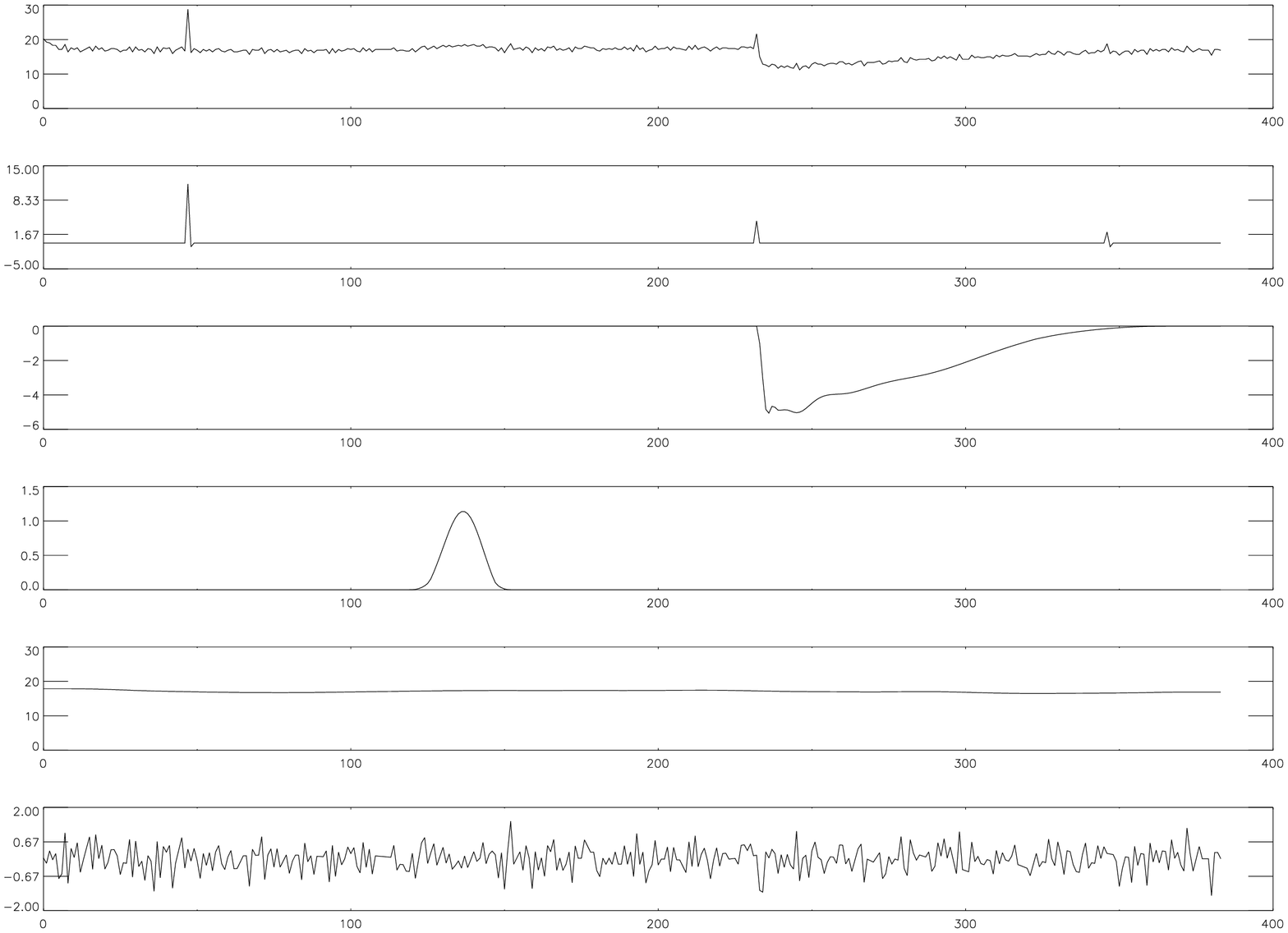,height=8.0cm,width=8.0cm}}
\caption{\em These plots show the decomposition of the signal
measured by a given pixel of the camera as a function of the number of
readouts (Starck $\et$ 1998). From top to bottom: the first plot
presents the typical time behaviour of a given ISOCAM pixel. One can
see from left to right: a "normal" glitch, a faint source which
extends over $\simeq$20 readouts, a "dipper", with long trough, and a
"normal" glitch again, at the end of the trough. The second plot,
shows the high frequency signal corresponding to the "normal" portion
of the glitches. The third plot shows the reconstructed signal of the
dipper's trough, after combining the information from several
frequencies or scales of the signal. The fourth plot presents a
gaussian at the position of the source with the extent of the source
(we do not extract the signal at this frequency since it contains
sources, but this plot is only intended to indicate the position of
the source). The fifth plot shows the low frequency part of the
signal, i.e. its baseline.  The last plot presents the residual of the
signal after extraction of the glitches and baseline. The source was
substracted to show only the gaussian noise residual, due to photon
and readout noise.  Very faint sources are not visible on the time
history of a given pixel and only appear after co-adding the residual
of all pixels which pointed towards the same sky direction on a final
raster map.}
\label{FIGURE:deconv}  
\end{figure}

ISOCAM data are subject to standard gaussian noise (photon and readout
noises) and to errors associated with the flat-fielding and dark
current substraction. But the main limitation of ISOCAM deep surveys
comes from its thick and cold pixels:
\begin{itemize}
\item because they are {\bf thick}, ISOCAM pixel detectors are very
sensitive to cosmic ray impacts (4.5 pixels receive a glitch per
second). The behaviour of these glitches can be divided into three
families:
\begin{itemize}
\item "normal glitches": the more common ones, which correspond to
electrons and last only one or two readouts. They are easily removed
with a median filtering (the combination of several scales for the
median allows the best correction).
\item "faders": these glitches as well as the following ones are
probably associated with protons and alpha particles. They induce
positive peaks in the detector response, which can last several
readouts. Since ISOCAM is best used in the raster mode, a real
source will look like these glitches, i.e. a positive response over
the number of readouts spent on a given position of the sky.
\item "dippers": some glitches are followed by a trough extended
over more than one hundred readouts.
\end{itemize}
We have developed a tool based on multi-resolution wavelet transform
which finds and removes the "faders" and "dippers". This tool, named
PRETI, for Pattern REcognition Technique for Isocam data (Starck $\et$
 1998), looks for the signal above the n-sigma level at different
frequencies or scales in the wavelet transform of the original
signal. Being extended, these glitches will be detected over several
successive scales as typical patterns. Using their wavelet
coefficients, one can substract a smooth function corresponding to the
glitch in the raw signal and use the output signal in the co-addition
of all pixels of the raster that fall on the same sky position (see
figure~\ref{FIGURE:deconv}).
\item ISOCAM pixels are {\bf cold}, so that electrons move very slowly
within them and therefore induce a transient behaviour: a pixel will
take several hundred readouts to stabilize when moving from the
background to the position of a source on the sky and inversely.
Because of time limitation, one is therefore limited to non stabilized
signals which results in an uncertainty on the photometry.  This
uncertainty is strongly reduced by the partial correction of the
transient behaviour and by the use of simulations to define a
statistical distribution of measured fluxes for any given input flux.
The final uncertainty on ISOCAM deep survey fluxes is on the order of
20 $\%$.
\end{itemize}

In order to facilitate the separation of sources from cosmic
ray impacts, ISOCAM surveys were performed using the raster
mode with a redundancy (number of different pixels falling
successively on a given sky position) ranging from 2 for the
shallowest survey (ELAIS) to 88 for the deepest surveys (Marano
Field Ultra-Deep survey; 64 for the HDF field).

\subsection{Simulations}
\label{SECTION:simulations}

\begin{figure}[!ht]
\centerline{\psfig{file=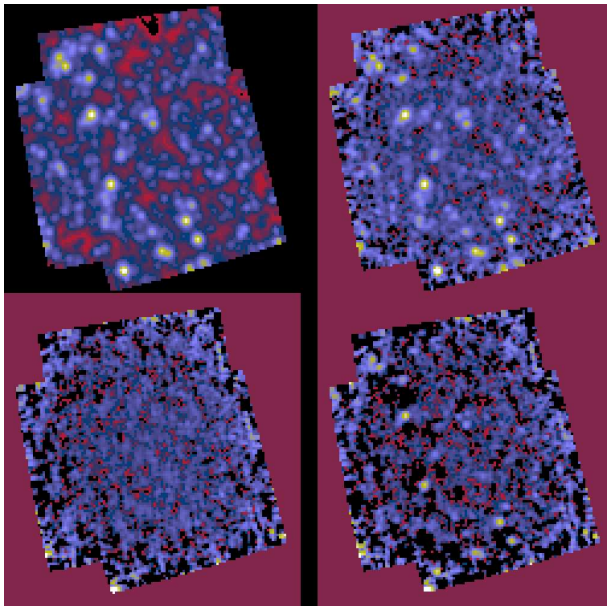,width=8.0cm,bbllx=1,bblly=1,bburx=360,bbury=360}}
\caption{\em simulated ISOCAM-HDF: from upper-left to bottom-right.
The first image shows the distribution of sources expected from the
model with strong evolution of Franceschini et al. 1994, where the
number of galaxies per unit luminosity per unit redshift is given by:
N(L,z)=N(L,z=0)$\times$exp($\kappa \tau$) where $\kappa$ is the
evolution parameter ($\kappa$=3)and $\tau$ is the lookback time=
$(t_0-t)/t_0$. Fluxes are ranging from 0.1 $\mu$Jy to 1 mJy. Only the
PSF is included and there is no noise associated with this ``ideal''
image.  In the second image, the ISOCAM gaussian noise resulting from
the readout and photon noise was added to the previous image.  The
last image presents the final image that ISOCAM would see including
the noise introduced by cosmic ray impacts, shown in the third image.
}
\label{FIGURE:simul}
\end{figure}

We performed simulations in order to quantify the sensitivity limit
(minimum detected flux, below the completeness limit), the
completeness limit (flux above which all sources are detected) and the
photometric accuracy. These simulations were performed with real
datasets (in order to simulate realistic glitches) in which we
introduced fake sources including the PSF and their modeled transient:
a long staring observation (more than 500 readouts) was used to
simulate a raster observation. Real sources would extend over the
whole observation and would be removed as a low frequency component of
the signal while in a raster observation sources only remain during
the number of readouts spent on a given position of the
sky. Figure~\ref{FIGURE:simul} shows as an example a simulation of
what an HDF-like ISOCAM deep survey would look like at 15 $\mu$m if
the source distribution followed the model developed by Franceschini
$\et$ (1991) with strong evolution ($\kappa = 3$). The next section
will present the real dataset to be compared with these simulations.

Finally, we also developed two independent data analysis techniques at
Saclay (PRETI, Starck $\et$ 1998) and Orsay (Tripple beam-switch technique,
see D\'esert $\et$ 1998), in order to test the confidence level
of the results. We find an agreement at the 20 $\%$ level
corresponding to the photometric error of both techniques, an
astrometric error close to the pixel size and compatible source
lists. In the case of the Hubble Deep Field, we compared our results
to the new results of the team of M.Rowan-Robinson (Rowan-Robinson $\et$
 1997, and references therein), which differ from their first analysis
in the photometry at 15 $\mu$m, because the techniques on each side
were improved with time (see Aussel $\et$ 1998).

\subsection{RESULTS}
\label{SECTION:results}

\begin{figure}[!ht]
\centerline{\psfig{file=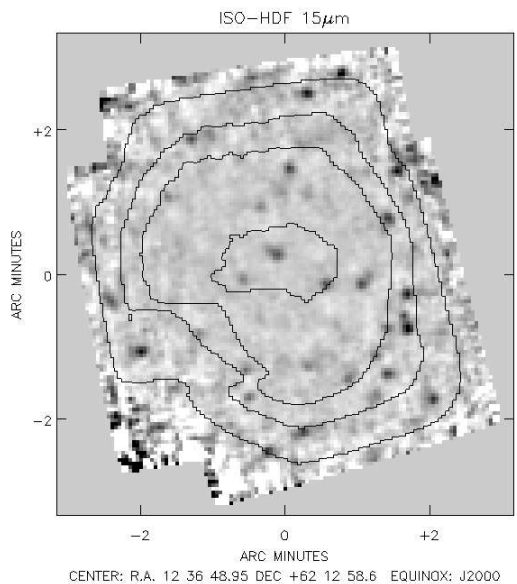,width=8.0cm,bbllx=54,bblly=362,bburx=413,bbury=721}}
\caption{\em ISOCAM 15 $\mu$m HDF image from Aussel et
al. (1998). Compare this image to the simulation shown in the previous
image. The excess of detections found in the real image reflects the
high evolution paramater needed to reproduce the number counts below
($\kappa$= 6 instead of 3).}
\label{FIGURE:hdf_lw3}
\end{figure}

\begin{figure}[!ht]
\centerline{\psfig{file=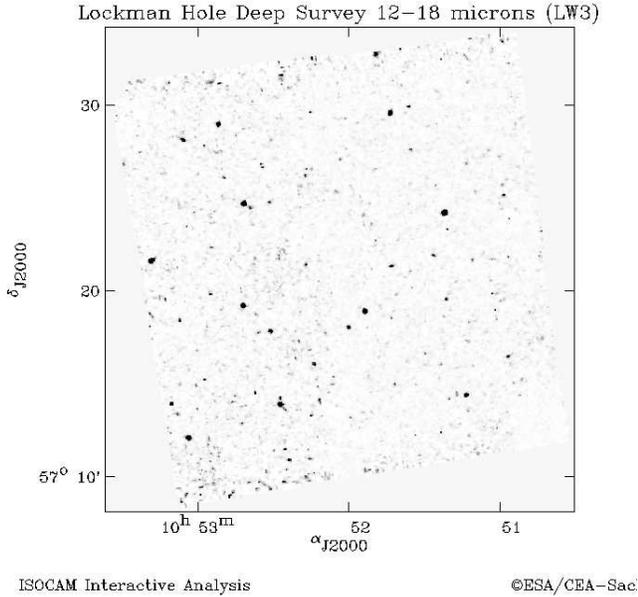,width=8.0cm,bbllx=54,bblly=362,bburx=413,bbury=721}}
\caption{\em ISOCAM 15 $\mu$m Lockman Hole Deep Survey image from Elbaz et al. in prep.}
\label{FIGURE:ds_lw3}
\end{figure}

Source detection was performed on the output images of PRETI
(see figure~\ref{FIGURE:hdf_lw3} $\&$ figure~\ref{FIGURE:ds_lw3} for the ISOCAM HDF
and Lockman Hole Deep Survey fields) using a spatial multi-scale
wavelet transform. Hence multi-scale transforms were performed at two
stages of the data analysis: first, on the temporal history of the
detector pixels in order to remove glitches, then on the final
co-added image for the source detection. The photometry of the Tripple
Beam-Switch technique from D\'esert $\et$ (1998) and PRETI are consistent at
the 20 $\%$ level.

\begin{figure}[!ht]
\centerline{\psfig{file=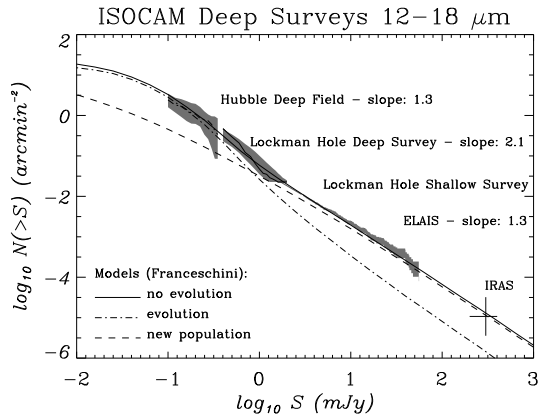,width=8.0cm,bbllx=54,bblly=362,bburx=556,bbury=720}}
\caption{\em log N($\>$S)- Log S: number of objects, N, detected above
a given flux level, S(mJy), with ISOCAM in the 15 $\mu$m broad band
filter LW3. The data are consistent although they come from 5
different origins. From high fluxes to low fluxes one finds:
ISOCAM-HDF (from Aussel et al. 1998), Lockman Hole deep survey (Elbaz
et al., in prep), Lockman Hole shallow survey (Elbaz et al. in prep), ELAIS
(preliminaty results from the ELAIS consortium, P.I. M.Rowan-Robinson,
Oliver et al. 1998), IRAS 12 $\mu$m (Spinoglio et al. 1995). Dashed
line: no evolution, normalized to IRAS and including PAH
features. Dashed-dotted line: strongly evolving population in
N(L,z)=N(L,z=0)$\times$exp($\kappa \tau$) with $\kappa$=6. Plain line:
sum of the no evolution $+$ strongly evolving population. The strongly
evolving population becomes dominant below 1 mJy}
\label{FIGURE:logN}
\end{figure}

Several hundreds of simulated images were done in order to estimate
from a Monte Carlo statistical approach the limits of our results in
terms of photometric accuracy and completeness. We have included this
information in the Log N-Log S curve (figure~\ref{FIGURE:logN}):
error bars on the fluxes and the flux range where completeness is
achieved result directly from the Monte Carlo simulations.

\begin{figure}[!ht]
\centerline{\psfig{file=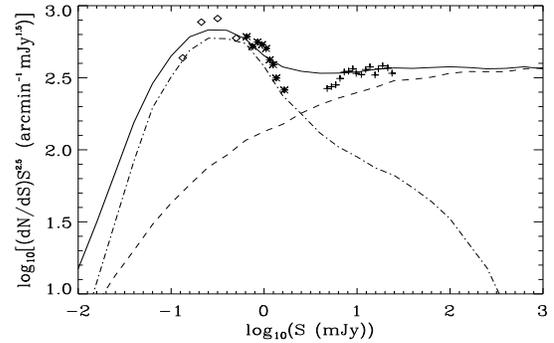,width=8.0cm,height=5.0cm}}
\caption{\em euclidean normalized differential counts: ISOCAM-HDF
(diamonds), Lockman Hole deep survey (stars), ELAIS (crosses). The
models are the same as in the log N-log S plot.}
\label{FIGURE:diff}
\end{figure}

The number counts are presented here in the two classical forms: the
log N-log S (figure~\ref{FIGURE:logN}), gives the number of objects
which flux is higher than a given flux limit S(mJy), and the
differential counts normalized to euclidean
(figure~\ref{FIGURE:diff}). No evolution would give a slope of 3/2 in
the Log N-log S in a euclidean space (2.5 in the differential counts,
by derivation). In a relativistic universe the slope of
the no evolution curve is even lower than that as shown by the
no-evolution curve from Franceschini (1998) overplotted on
figure~\ref{FIGURE:logN}.

\begin{figure}[!ht]
\centerline{\psfig{file=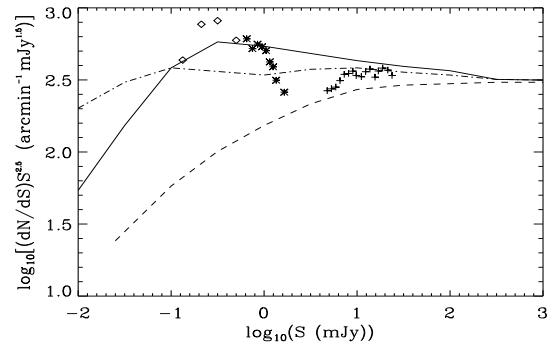,width=8.0cm,height=5.0cm}}
\caption{\em euclidean normalized differential counts: same as
above. Models of Xu et al. (1998) normalized to the same IRAS point as in figure \ref{FIGURE:diff} . Plain line: pure
luminosity evolution, $\rho \propto (1+z)^{4}$. Dashed-dotted line:
pure density evolution, $\rho \propto (1+z)^{3}$. Dashed line: no evolution.}
\label{FIGURE:diff_xu}
\end{figure}


\subsection{Interpretation}
\label{SECTION:counts}

For the first time, we present number counts over 4 orders of
magnitude ranging from 100 $\mu$Jy, in the ISOCAM-HDF (Aussel $\et$
1998), to 50 mJy, in the ELAIS fields (preliminary results from the
ELAIS consortium, P.I. M.Rowan-Robinson, Oliver et al. 1998) and up to
300 mJy, including the IRAS 12 $\mu$m counts. It can be seen on both
the log N-log S (figure~\ref{FIGURE:logN}) and the differential counts
(figure~\ref{FIGURE:diff}) that above the 1 mJy level, the counts
follow the no-evolution model, while below this flux level, a rapid
rise appears (clearly seen on the differential counts plot). Such a
trend cannot be reproduced either by the PAHs features, which effect
on the K-correction is too weak, or by a continuous evolution such as~:
\begin{enumerate}
\item pure luminosity evolution as described in Franceschini $\et$
(1991): N(L,z)=N(L,0) $\times$ exp($\kappa \tau$) where $\kappa$ is
the evolution parameter and $\tau= (t_0-t)/t_0$ is the lookback
time. No value of $\kappa$ allows to fit the data, except if we
consider a separate population of objects which becomes dominant below
the 1 mJy level, i.e. above z$\simeq$0.4.
\item pure luminosity evolution used by Pearson $\&$
Rowan-Robinson (1996) to fit the IRAS 60 $\mu$m number counts,
proportional to $(1+z)^3$ and calculated by Xu $\et$ (1998) at 15
$\mu$m (figure~\ref{FIGURE:diff_xu}).
\item the pure density evolution, in $(1+z)^4$, from Xu $\et$ (1998,
figure~\ref{FIGURE:diff_xu}).
\end{enumerate}
\begin{figure}[!ht]
\centerline{\psfig{file=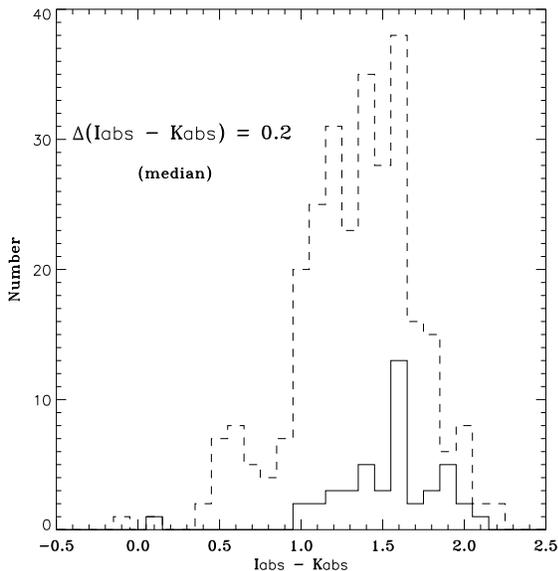,width=8.0cm,bbllx=56,bblly=362,bburx=478,bbury=785}}
\caption{\em Histogram of the I-K rest-frame colours of all
HDF$+$Flanking fields galaxies (dashed line) and HDF$+$Flanking fields
galaxies detected at 15 $\mu$m above 100 $\mu$Jy. The rest-frame
colours were calculated using the HDF and Flanking Field catalog of
Barger $\et$ (1998b) with a method similar to the one used by Lily $\et$ 
(1995). The I band K-correction was computed by us using the
Coleman $\et$ 1980 SED and the K band K-correction is from Cowie $\et$
(1994). The difference in the median colours is only 0.2 dex.}
\label{FIGURE:colors}
\end{figure}
Extreme evolution needs to be added to the no-evolution curve, which
could be associated with a galaxy population dominating the counts
below 1 mJy, i.e. above z=0.4 (in this model). The evolution
parameter, $\kappa$, required for this ``population'' is $\kappa$=6,
while it was 3, in the original model with strong evolution of
Franceschini $\et$ (1991). This rapid increase of the number counts
below 1 mJy, corresponds to a high flux density in the mid-IR going
towards faint fluxes. Integrating the number counts over the whole
flux range one finds that about $2\times10^{-9}~W.m^{-2}.sr^{-1}$ is
emitted at 15 $\mu$m and $3\times10^{-9}~W.m^{-2}.sr^{-1}$ when
extrapolating the counts assuming no further increase in the
slope. This corresponds to 30-45 $\%$ of the cosmic background seen
with the HST in the I-band ($7\times10^{-9}
~W.m^{-2}.sr^{-1}$). Locally about one third of the energy is radiated
in the infrared (Soifer $\&$ Neugebauer 1991) and mainly in the FIR
since the FIR over MIR ratio is typically 3 for normal galaxies and
stronger than 10 for starbursting galaxies. The cosmic infrared
background is therefore larger than the optical background when
integrating over a larger redshift range than IRAS, which implies that
going towards higher redshifts:
\begin{itemize}
\item dust obscuration is stronger at higher redshift. 
\item a few bright infrared objects make most of the cosmic background,
since the few detections at 15 $\mu$m in the HDF radiate more energy
,when integrated over the whole infrared range, than the large number
of optically detected galaxies.
\end{itemize}
\begin{figure}[!ht]
\centerline{\psfig{file=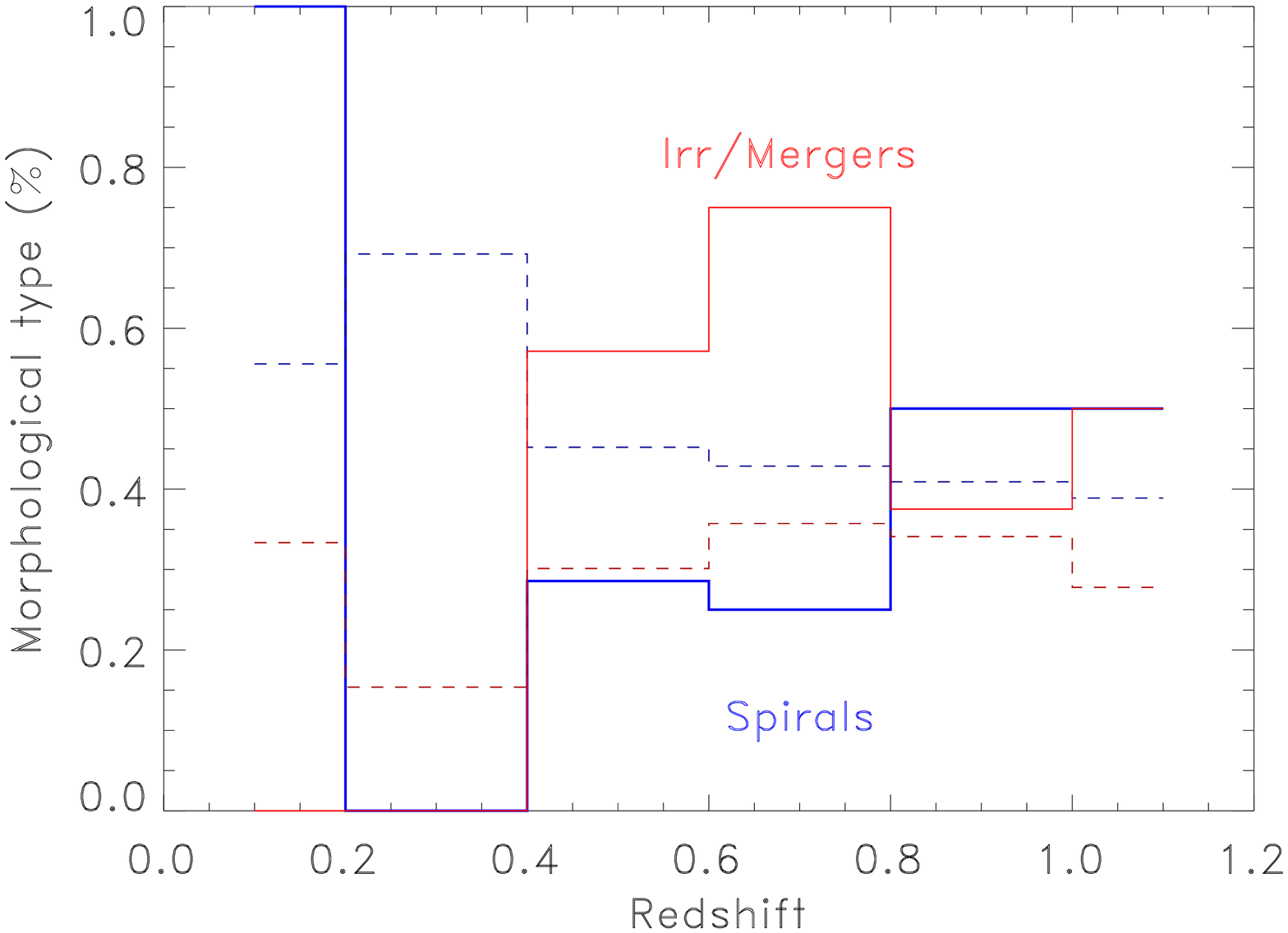,width=8.0cm,bbllx=1,bblly=1,bburx=685,bbury=505}}
\caption{\em Fraction of Peculiar versus Spiral galaxies in the
HDF$+$Flanking fields for galaxies detected at 15 $\mu$m (plain lines)
and all galaxies (dashed lines). Above z$\simeq$0.4, peculiar galaxies
dominate the distribution of galaxies detected with ISOCAM while
optically selected galaxies are mainly spirals. This population of
peculiar galaxies may produce the rapid evolution below the 1 mJy
level found in the 15 $\mu$m number counts.}
\label{FIGURE:morpho}
\end{figure}
This result agrees with SCUBA data (Barger $\et$ 1998a, Hughes $\et$
1998) and with the CIRB estimates (see Puget $\et$, these
proceedings). Additional conclusions can be reached only after
identification of the sources, and multi-wavelength and redshift
measurements. We are currently gathering such data for the sources in
the Lockman Hole. In the HDF and in one of the CFRS fields, surveyed
with ISOCAM by Flores $\et$ (1998), such data are available. We
summarize some of the main conclusions obtained from this source by
source comparison over $\simeq$ 60 sources within the HDF$+$flanking
fields with redshifts in the range 0.4-1.3, as well as in the CFRS~:
\begin{itemize}
\item at z$>$0.5, peculiar, mergers and post-starburst galaxies are
much more frequent in the ISO base than in the optical base (Flores
$\et$ 1998, Elbaz $\et$ (in prep.), see figure~\ref{FIGURE:morpho}). It is likely
that the additional population required to account for the number
counts is made of such galaxies.
\item Flores $\et$ (1998) attempt to fit the multi-wavelength data
over 10 sources for which radio data are available. In this way, they
estimate that the universal star formation at z$\simeq$1 is enhanced
by a factor 2.3 with respect to previously estimated UV fluxes.
\end{itemize}

Using available data on the HDF, especially optical and NIR data from
Barger $\et$ (1998b), we find that galaxies detected above 100 $\mu$Jy
(completeness limit reached in the HDF, Aussel $\et$ 1998) do not show
any clear signature in their optical colours (see
figure~\ref{FIGURE:colors}, Elbaz $\et$ {\it in prep}). The I-K colour of
ISOCAM detected galaxies is only 0.2 dex redder than all HDF galaxies.


\subsection{CONCLUSION}
\label{SECTION:conclusion}

We showed that even with very sensitive telescopes like the HST,
optical observations need to be combined with infrared imaging in
order to select the major events of star formation in the universe.
The sensitivity required for the NGST to detect all HDF optical
galaxies, i.e. 2$\times10^6$ galaxies per square segree, at 15 $\mu$m
would be about 0.02 $\mu$Jy if the slope flattens as shown by the
model on figure~\ref{FIGURE:logN} and 0.3 $\mu$Jy if a moderate
evolution with a slope close to 1.3 continues below the 100 $\mu$Jy
level. For a large fraction of these galaxies, the NGST would resolve
the regions of star formation in actively star forming galaxies and
change our view on galaxy evolution without suffering from dust
obscuration.



\end{document}